%% file: known_ENR_journal.tex
\documentclass[journal,draftclsnofoot,onecolumn,12pt,twoside]{IEEEtran}
\IEEEoverridecommandlockouts

\usepackage{amsthm}

\newtheorem{thm}{Theorem}
\newtheorem{prop}{Proposition}

\theoremstyle{definition}
\newtheorem*{defn*}{Definition}
\newtheorem*{scheme*}{Scheme}

\theoremstyle{remark}
\newtheorem{remark}{Remark}

\input{defs}

\usepackage{ifthen}
\usetikzlibrary{circuits.ee.IEC}
\usepackage{circuitikz}

\usepackage{mathtools}

\mathtoolsset{showonlyrefs=true}

\usepackage{MnSymbol}

\usepackage{comment}
\newcommand{\ignore}[1]{}

\newcommand{\VersionLength}{short}

\providecommand{\ver}{\ifthenelse{\equal{\VersionLength}{long}}}
\providecommand{\nver}{\ifthenelse{\equal{\VersionLength}{short}}}

\providecommand{\figref}[1]{Fig.~\ref{#1}}
\providecommand{\secref}[1]{Sec.~\ref{#1}}

\providecommand{\ColumnNum}{1}
\newcommand{\col}{\ifthenelse{\equal{\ColumnNum}{1}}}

\begin{document}
\title{Energy-limited Joint Source--Channel Coding via Analog Pulse Position Modulation}

\author{Omri Lev and Anatoly Khina
    \thanks{This work was supported by 
    the \textsc{Israel Science Foundation} (grant No.\ 2077/20).
    The work of O.~Lev was further supported by the Yitzhak and Chaya Weinstein Research Institute  for Signal Processing. 
    The work of A.~Khina was further supported by the WIN Consortium through the Israel Ministry of Economy and Industry.
    This paper will be presented in part at the
    IEEE Workshop on Information Theory, Oct.\ 2021.}
    \thanks{The authors are with the Department of Electrical Engineering--Systems, Tel Aviv University, Tel Aviv, Israel~6997801 (e-mails: \texttt{omrilev@mail.tau.ac.il,anatolyk@eng.tau.ac.il}).}
}
\maketitle


\begin{abstract}
    We study the problem of transmitting a source sample with minimum distortion over an infinite-bandwidth additive white Gaussian noise channel under an energy constraint. To that end, we construct a joint source--channel coding scheme using analog pulse position modulation (PPM) and bound its quadratic distortion. We show that this scheme outperforms existing techniques since its quadratic distortion attains both the exponential and polynomial decay orders of Burnashev's outer bound. We supplement our theoretical results with numerical simulations and comparisons to existing schemes.
\end{abstract}


\allowdisplaybreaks

\section{Introduction}
\label{s:intro}

Recent developments in distributed sensor arrays and the internet of things raise the need for communicating a small number of measurements with small distortion over wireless media across large time spans and/or large bandwidth with limited energy.
Indeed, since such sensors have limited battery, due to environmental (in the case of energy harvesting) or replenishing limitations, the required transmission solutions need to be economical in terms of energy; on the other hand, since the measured signal needs to be transmitted with low frequency (a single measurement at a time), 
the time/frequency used by the sensor for the transmission of each measurement may be quite large.

This problem may be conveniently modeled as the classical setup of conveying $k$ independent and identically distributed samples of a source over a continuous-time additive white Gaussian noise (AWGN) channel under an energy constraint per source sample. 

In the limit of a large source blocklength, $k \to \infty$, 
the optimal performance is known and is dictated by the celebrated source--channel separation principle \cite[Th.~10.4.1]{CoverBook2Edition}, \cite[Ch.~3.9]{ElGamalKimBook}. 
For a memoryless Gaussian source and a quadratic distortion measure, the minimal (optimal) achievable distortion $D$ is given by 
\begin{align}
    D = \sigma_x^2 \cdot \e^{-2\ENR},
\label{eq:SeparationBound}
\end{align}
where $\ENR$ denotes the energy-to-noise ratio (ENR) over the channel, and $\sigma_x^2$ is the source variance.
For other continuous memoryless sources, the optimal distortion is bounded as \cite[Prob.~10.8, Th.~10.4.1]{CoverBook2Edition}, \cite[Prob.~3.18, Ch.~3.9]{ElGamalKimBook}
\begin{align}
    \frac{\e^{2h(x)}}{2\pi\e} \cdot \e^{-2\ENR} \leq D \leq \sigma_x^2 \cdot \e^{-2\ENR} ,
\end{align}
where the lower bound stems from Shannon's lower bound \cite{Shannon59:RDF}, the upper bound holds since a Gaussian source is the ``least compressable'' source with a given variance under a quadratic distortion measure, and $h(x)$ denotes the differential entropy (in nats) of the source $x$ \cite[Ch.~8]{CoverBook2Edition}, \cite[Ch.~2.2]{ElGamalKimBook}.

While the optimal performance is known in the limit of large blocklength,
determining it becomes much more challenging when the blocklength $k$ is finite.

For a scalar source ($k = 1$), both lower and upper bounds on the achievable distortion have been constructed.
Lower (impossibility) bounds on the distortion have been devised in \cite{Cohn_Phd,BurnashevInfiniteBandwidthExponent,BurnashevInfiniteBandwidthExponent1}, 
with Burnashev \cite{BurnashevInfiniteBandwidthExponent} providing the tightest lower bound on the distortion, for large ENRs, of 
\begin{align} 
    D \geq K_1 \cdot \ENR^{-K_2} \cdot \e^{-\frac{\ENR}{3}} \{1 + o(1)\}
\label{eq:D:Burnashev}
\end{align} 
for some positive constants $K_1$ and $K_2$ (which are not known explicitly),
where $o(1) \to 0$ for $\ENR \to \infty$.
Separation-based schemes that attain 
\begin{align} 
    D \leq K \e^{-\frac{\ENR}{3}} \{ 1 + o(1)\},
\label{eq:D:separation}
\end{align} 
where again $o(1) \to 0$ for $\ENR \to \infty$,
have been constructed in \cite{BurnashevInfiniteBandwidthExponent1,SeparationInfBW_Abdallah,TuncelInfinitedBW_SeparationCompanding:Journal}. These schemes employ scalar quantization in conjunction with orthogonal signaling---e.g., pulse position modulation (PPM)---that is known to be capacity-achieving in the power/energy limited regime \cite[Ch.~8]{WozencraftJacobsBook}, \cite[Ch.~8]{GallagerBook1968}, \cite[Ch.~2.5]{ViterbiOmuraBook}. In particular, Sevin\c{c} and Tuncel~\cite{TuncelInfinitedBW_SeparationCompanding:Journal} optimized the constant $K$ in \eqref{eq:D:separation} by appealing to high-resolution quantization (\textit{a la} Bennett's asymptotic quantization \cite{Bennett48,PanterDite:FLC}, \cite[Ch.~5.6]{GershoGrayBook}).

However, by comparing \eqref{eq:D:Burnashev} with \eqref{eq:D:separation}, 
one observes a gap in the asymptotic behavior manifested by the polynomial factor $\ENR^{-K_2}$ that needs to be closed.

In this work, we construct a purely joint source--channel coding (JSCC) scheme for the setting of a scalar source and known ENR, in lieu of the available separation-based solutions: Instead of using quantization and digital PPM, we map the source sample into a shift of a rectangular pulse directly. We show that this scheme achieves Burnashev's outer bound \eqref{eq:D:Burnashev} for some positive $K_1$ and $K_2$, and is therefore strictly better than the hitherto available solutions \eqref{eq:D:separation} which fail to attain the additional polynomial improvement factor of $\ENR^{-K_2}$ of \eqref{eq:D:Burnashev}.

The rest of the paper is organized as follows. 
We 
introduce the notation that is used in this work in \secref{ss:notation}, 
and 
formulate the problem setup in \secref{s:Problem Statement}. 
We construct an analog PPM scheme and analyze its performance in \secref{s: AnalogPPM}; numerical comparison of the derived bounds to Monte Carlo simulations 
of the proposed scheme is presented in \secref{s:numeric}. 
We conclude the paper with a discussion and propose future research directions in \secref{s:Summary}.


\subsection{Notation}
\label{ss:notation}

$\nats$, $\reals$, $\reals_+$ denote the sets of the natural, real and the non-negative real numbers, respectively.
With some abuse of notation, 
we denote tuples (column vectors) by $a^{k} \triangleq \left(a_0, \ldots, a_{k-1} \right)^\dagger$ for $k \in \nats$, and their Euclidean norms---by $\norm{a^k} \triangleq \sqrt{\sum_{i=0}^{k-1} a_i^2}$, where $(\cdot)^\dagger$ denotes the transpose operation; distinguishing the former notation from the power operation applied to a scalar value will be clear from the context.
The complement of an event $A$ is denoted by $\bar{A}$. 
All logarithms are to the natural base and all rates are measured in nats.
$\E{\cdot}$ and $\CE{\cdot}{\cdot}$ 
denote the expectation and conditional expectation operations, respectively.

	
\section{Problem Statement}
\label{s:Problem Statement}

In this section, we formalize the JSCC setting that will be treated in this work.

\textit{Source.}
The source sample to be conveyed, $x \in \reals^k$, is distributed according to a known probability density function (p.d.f.) $f_x$ and has variance $\sigma_x^2$. We will consider two specific source distributions:
\begin{itemize}
\item 
    A continuous uniform distribution over 
    $[-0.5, 0.5]$.
\item 
    A standard Gaussian distribution.
\end{itemize}

\textit{Transmitter.}
Maps the source sample $x$ to a continuous input waveform $\left\{ s_{x}(t) \middle|  |t| \leq T/2 \right\}$ that is subject to an energy constraint:\footnote{The introduction of negative time instants yields a non-causal scheme. This scheme can be made causal by introducing a delay of size $T/2$. We use a symmetric transmission time around zero for convenience.}$^\text{,}$\footnote{The resulting average power $P$ is therefore equal to $P = E/T$.}
\begin{align}
\label{eq:InputEnergyConstraint}
    \int_{-\frac{T}{2}}^{\frac{T}{2}}\abs{s_{x}(t)}^2dt &\leq E, &\forall x \in \reals,
\end{align}
where $E$ denotes the transmit energy.

\textit{Channel.}
$s_{x}$ is transmitted over a continuous-time AWGN channel:
\begin{align}
 \label{eq:ChannelEq}
     r(t) &= s_{x}(t) + n(t), & t \in \left[-\frac{T}{2},\frac{T}{2}\right],
\end{align}
where $n$ is a continuous-time AWGN with two-sided spectral density $N_0/2$,  
and $r$ is the channel output signal.

\textit{Receiver.}
Receives the channel output signal $r$,
and constructs an estimate $\hx$ of $x$.

\textit{Distortion.}	
The average quadratic distortion 
between $x$ and $\hx$
is defined as
\begin{align}
\label{eq:Distortion}
    D \triangleq \frac{1}{k}\E{\norm{x^k - \hx^k}^{2}},
\end{align}	
and the corresponding signal-to-distortion ratio (SDR)---by
\begin{align}
\label{eq:SDR}
    \SDR \triangleq \frac{\E{x^2}}{D}.
\end{align}	

\textit{Regime.}
We concentrate on the energy-limited regime, \viz\ the channel input is not subject to a power or a bandwidth constraint, but rather to a per-symbol energy constraint. 

As specified in \eqref{eq:ChannelEq}, the channel input is subject to an energy constraint $E$, the capacity of which is equal to \cite[Ch.~9.3]{CoverBook2Edition} 
 \begin{align}
     \label{eq:ChannelCapacity_Lim}
     C = \ENR,
 \end{align}
where $\ENR \triangleq E/N_0$ is the ENR,
and the capacity is measured in nats;
note that the available bandwidth is unconstrained (\ie, infinite).


\section{Main Results}
\label{s: AnalogPPM}

In this section, 
we employ analog PPM and derive upper bounds on its distortion for uniform and Gaussian sources in Secs.~\ref{s:UpperBound_UniformPrior} and \ref{s:UpperBound_GaussianPrior}, respectively.
The performance of this scheme, hinges on the ability of the receiver to estimate the position of a transmitted pulse with known shape corrupted by AWGN:
Consider the problem statement of \secref{s:Problem Statement} 
with the following specific modulation:
\begin{align}
\label{eq:PPM:waveform}
    s_x(t) = \sqrt{E}\phi(t - x \Delta)
\end{align}
where $\phi$ is a predefined pulse with unit energy and $\Delta$ is a scaling parameter.

This fundamental problem received much attention over the years because of its importance in classical applications such as radar and in emerging applications such as the internet of things. Nonetheless, closed-form expressions for the optimal receiver and its distortion \eqref{eq:Distortion} remain an open problem, in general. 
Consequently, various bounds and approximations have been derived over the years; see e.g., \cite{ChazanZivZakai:ParamterEstimationBound:IT1975,WeissWeinstein:TOAbound,Zehavi:TOAbound,ZivZakai:ThresholdPPM_Rect,Merhav:StatPhys_ThresholdPPM},\cite[Ch.~8]{WozencraftJacobsBook} (and the reference therein).
Interestingly, the available results suggest that the shape of the pulse $\phi$ may have a big effect on the achievable performance as well as the distribution of $x$; specifically, a rectangular pulse is known to achieve good performance as we detail next.

Thus, 
we 
concentrate on the case of a rectangular pulse:\footnote{Clearly, the bandwidth of this pulse is infinite. By taking a large enough bandwidth $W$, one may approximate this pulse to an arbitrarily high precision and attain its performance within an arbitrarily small gap.} 
\begin{align}
    \label{eq:AnalogPPM:PulseShaping_RectPulse}
    \phi(t) &=
    \begin{cases}
        \sqrt{\frac{\beta}{\Delta}}, & \abs{t} \leq \frac{\Delta}{2\beta},\\
        0, & \mathrm{otherwise},
    \end{cases}
\end{align}    
for a parameter $\beta > 1$ which is sometimes referred to as \textit{effective dimensionality}. Clearly, $T = \Delta + \Delta/\beta$. 

The optimal receiver is the MMSE estimator $\hx$ of $x$ given the entire output signal:
\begin{align}
    \hx^\MMSE = \CE{x}{r}.
\label{eq:delay-estimate:MMSE}
\end{align}


\subsection{Upper Bound on the Distortion for a Uniform Source}
\label{s:UpperBound_UniformPrior}

We construct here an upper bound on the achievable distortion of the proposed analog PPM scheme for a uniform source. 
For the sake of analysis, we examine the performance of the (suboptimal) maximum a posteriori (MAP) estimator instead of the (optimal) MMSE estimator \eqref{eq:delay-estimate:MMSE}, which, for the case of a uniform source, reduces, to the maximum-likelihood (ML) estimator:
\begin{align}
\label{eq:MAP:uniform}
\begin{aligned}
    \hx^\MAP 
    &= \argmax_{\hx:\ \abs{\hx} \leq 0.5} \ \e^{-\frac{1}{N}\int_{-\infty}^\infty \left(r(t) - s_{\hx} (t) \right)^2dt} 
 \\ &= \argmin_{\hx:\ \abs{\hx} \leq 0.5} \int_{-\infty}^\infty \left(r(t) - s_{\hx} (t)\right)^2dt.
\end{aligned}
\end{align}

Since the waveform $s_x$ of \eqref{eq:PPM:waveform}
has energy $E$ for all $x$, the ML estimator reduces further to maximum correlation:
\begin{align}
\label{eq:AnalogPPM:Receiver_MaxCorr}
\begin{aligned} 
    \hx^\MAP &= \argmax_{\hx:\ \abs{\hx} \leq 0.5} \int_{-\infty}^\infty r(t) s_{\hx} (t) dt 
 \\ & = \argmax_{\hx:\ \abs{\hx} \leq 0.5} \int_{-\infty}^\infty r(t)\phi(t - \hx \Delta) dt 
 \\ &= \argmax_{\hx:\ \abs{\hx} \leq 0.5} R_{r,\phi}(\hx \Delta),
\end{aligned}
\end{align}
where 
\begin{subequations} 
\label{eq:correlations}
\noeqref{eq:Rphi}
\begin{align} 
\begin{aligned} 
    R_{r,\phi}(\hx \Delta) &\triangleq \int_{-\infty}^\infty r(t)\phi(t - \hx \Delta) dt
 \\ &= \sqrt{E} R_\phi \left( (x-\hx) \Delta \right) + \sqrt{\frac{\beta}{\Delta}} \int_{\hx \Delta - \frac{\Delta}{2\beta}}^{\hx \Delta + \frac{\Delta}{2\beta}} n(t) dt ,
\end{aligned}
\label{eq:Rr_phi}
\end{align} 
is 
the (empirical) cross-correlation function between 
$r$ and $\phi$ with lag (displacement)~$\hx \Delta$,
and 
\begin{align}
\begin{aligned} 
    R_{\phi} (\tau) &= \int_{-\infty}^\infty \phi(t) \phi(t - \tau) dt 
 \\ &= 
    \begin{cases}
        1 - \frac{|\tau|}{\frac{\Delta}{\beta}}, & |\tau| \leq \frac{\Delta}{\beta}
     \\ 0, & \mathrm{otherwise}
    \end{cases}
\end{aligned}
\label{eq:Rphi}
\end{align}
\end{subequations}
is the autocorrelation function of $\phi$ with lag $\tau$.

The next proposition bounds from above the distortion of this receiver.

\begin{prop}
\label{prop:UpperBound_UniformPrior}
    The distortion of the MAP estimator \eqref{eq:AnalogPPM:Receiver_MaxCorr} of a scalar source that is uniformly distributed over a unit interval,  transmitted using analog PPM with a rectangular pulse is bounded from above by
    \begin{align}
        \label{eq:UpperBounf_UniformPrior_Explicit}
        D \leq D_S + P_L D_L,
    \end{align}
    where
    \begin{align}
        D_S &\triangleq \tilde D_S \cdot \left( 1 + \frac{16}{13} \sqrt{\frac{\ENR}{2}}\cdot e^{-\frac{\ENR}{4}} \right), \\
        D_L &\triangleq \frac{1}{6}\left(1 + \frac{2}{\beta} + \frac{4}{\beta^2}\right),\\
        P_L &\triangleq \tP_L \cdot  \left(1 + \sqrt{\frac{3}{4}}\frac{\e^{-\frac{\ENR}{6}}}{\sqrt{\ENR}} + 4\sqrt{\frac{\pi}{\ENR}}\right),
    \end{align}
    are upper bounds on the ``small-error'' distortion (when the error is less than or equal to $1/\beta$), ``large-error'' distortion, and the probability of a large error, respectively, 
    and
    \begin{align}
        \tilde D_S &\triangleq  \frac{13/8}{(\beta\ENR)^2} ,
     \\ \tilde P_L &\triangleq \frac{\beta \sqrt{\ENR} \e^{-\frac{\ENR}{2}}}{16\sqrt{\pi}} . 
    \end{align}
    In particular, in the limit of large $\ENR$, and $\beta$ that increases monotonically with $\ENR$,
    \begin{align}
    \label{eq:UpperBounf_UniformPrior_Explicit2}
        D \leq (\tilde D_S + \tilde P_L D_L)\left\{ 1 + o(1) \right\}
    \end{align}
    where $o(1) \to 0$ in the limit of $\ENR \to \infty$.
\end{prop}

\quad \textit{Proof:}
    Denote the estimation error by $\eps \triangleq x - \hx$. 
    Then, by the law of total expectation:
    \begin{subequations}
    \label{eq:UpperBound}
    \noeqref{eq:UpperBound_TotalExpec}
    \begin{align}
        \E{\eps^2} &= P\left(\abs{\eps} \leq \frac{1}{\beta}\right)\CE{\eps^2}{\abs{\eps} \leq \frac{1}{\beta}} 
    \col{}{\\* & \phantom{\,\, \leq \CE{\eps^2}{\abs{\eps} \leq \frac{1}{\beta}}}}
        + P\left(\abs{\eps} > \frac{1}{\beta}\right)\CE{\eps^2}{\abs{\eps} > \frac{1}{\beta}}     
    \label{eq:UpperBound_TotalExpec}
     \\ &\leq \CE{\eps^2}{\abs{\eps} \leq \frac{1}{\beta}} 
        + P\left(\abs{\eps} > \frac{1}{\beta}\right)\CE{\eps^2}{\abs{\eps} > \frac{1}{\beta}} .\ \ 
        \label{eq:UpperBound_TotalExpec_Mid}
    \end{align}
    \end{subequations}

    We now bound the terms in \eqref{eq:UpperBound_TotalExpec_Mid} by $D_S, P_L$ and $D_L$. 

    By \cite[Eq.~6]{Zehavi:TOAbound}--- 
    \begin{align}
        \label{eq:Zehavi:TOA_UB}
         \CE{\eps^2}{\abs{\eps} \leq \frac{1}{\beta}}  \leq D_S,
    \end{align}
  while by \cite[Eq.~15]{ZivZakai:ThresholdPPM_Rect}---
    \begin{align}
    \label{eq:ZZ:Th_UB}
     P\left(\abs{\eps} > \frac{1}{\beta}\right) &\leq P_L .
     \end{align}
     
     To bound the remaining term in \eqref{eq:UpperBound_TotalExpec_Mid}, 
     note that for $|\eps| > \frac{1}{2\beta}$, or equivalently $|(x - \hx)\Delta| > \frac{\Delta}{2\beta}$, 
     the autocorrelation term $R_\phi((x-\hx)\Delta)$ in \eqref{eq:Rr_phi} is nullified and only a noise term remains. 
%
    Thus, using \eqref{eq:AnalogPPM:Receiver_MaxCorr}, \eqref{eq:correlations}, the stationarity of the noise process $n$ and symmetry, we arrive at 
     \begin{align}
         &\!\!\!\!\!\!
         \CE{\eps^2}{\abs{\eps} > \frac{1}{\beta}} 
         = \CE{\left( x - \argmax_{\hx:\ \abs{\hx} \leq \frac{1}{2}} R_{r,\phi}(\hx \Delta) \right)^2}{\abs{\eps} > \frac{1}{\beta}}
      \\ &\!\!\!\!\!\!
      = \CE{\left( x - \sqrt{\frac{\beta}{\Delta}} \argmax_{\hx:\ \abs{\hx} \leq \frac{1}{2}} \int_{\hx \Delta - \frac{\Delta}{2\beta}}^{\hx \Delta + \frac{\Delta}{2\beta}} n(t) dt \right)^2}{\abs{x-\hx} > \frac{1}{\beta}} \quad
    \label{eq:LargeError_Uniform}
      \\ &\!\!\!\!\!\!
      = 2 \int_\frac{1}{\beta}^\frac{1}{2} e^2 \frac{de}{\frac{1}{2} - \frac{1}{\beta}} 
      \\ &\!\!\!\!\!\!
      = D_L . \tag*{\IEEEQED}
    \end{align}

By optimizing over $\beta$, we are able to derive the following upper bound on the achievable distortion. 

\begin{thm}
\label{thm:analog-PPM:knownENR:uniform:achievable}
    The achievable distortion of a uniform scalar source transmitted over an energy-limited channel with a known ENR is bounded from above as 
    \begin{align}
    \label{eq:KnownENR:RectPulse_UpperBoundOptim_Final}
        D &\leq 0.072\, \e^{-\frac{\ENR}{3}}\cdot \left(\ENR\right)^{-\frac{1}{3}} \cdot \left\{1 + o(1)\right\}
        ,
    \end{align}
    where 
    $o(1) \to 0$ as $\ENR \to \infty$.
\end{thm} 

\begin{IEEEproof}
    Setting $\beta = \left(312\sqrt{\pi}\right)^{\frac{1}{3}}\left(\ENR\right)^{-\frac{5}{6}} \e^{\frac{\ENR}{6}}$
    in \eqref{eq:UpperBounf_UniformPrior_Explicit2} of \propref{prop:UpperBound_UniformPrior} yields \eqref{eq:KnownENR:RectPulse_UpperBoundOptim_Final}.
\end{IEEEproof}

Thus, using the analog PPM scheme, we are able to meet Burnashev's asymptotic outer bound \eqref{eq:D:Burnashev} with $K_1 = 0.072$ and $K_2 = 1/3$. 
Therefore, in addition to attaining the optimal exponential decay with the ENR, it achieves also the next-order polynomial decay with the ENR. 
This offers an improvement over the asymptotic performance \eqref{eq:D:separation} of the separation-based scheme of Sevin\c{c} and Tuncel \cite{TuncelInfinitedBW_SeparationCompanding:Journal}, 
which, despite attaining the best exponential decay of $1/3$, does not attain the additional polynomial decay with the ENR.

\begin{remark}
\label{rem:AnaolgPPM:quadratic-improvement}
    For a fixed $\beta$, the distortion improves quadratically with the $\ENR$. This behavior will proof useful in the next section, where we address the unknown-ENR regime, and holds also for a Gaussian source as will become evident in \secref{s:UpperBound_GaussianPrior}.
\end{remark}

\begin{remark}
\label{rem:EqSDR_PPM}
    The bound of \thmref{thm:analog-PPM:knownENR:uniform:achievable} can be equivalently represented in terms of the SDR as 
    \begin{align}
        \label{eq:PPM_SDR}
        \SDR = \frac{1}{12D} \geq \kappa \e^{\frac{\ENR}{3}}\left(\ENR\right)^{\frac{1}{3}}\cdot\left\{1 + o(1)\right\} 
    \end{align}
    where $\kappa \triangleq \frac{1}{12K} \geq 1.1518$.
    By interpreting the SDR as the SNR of an effective channel with additive uncorrelated noise from the source $x$ to the estimator $\hx$. 
\end{remark}


\subsection{Upper Bound on the Distortion for a Gaussian Source}
\label{s:UpperBound_GaussianPrior}

We now construct an upper bound on the achievable distortion of the proposed analog PPM scheme for a Gaussian source. 

Again, for the sake of analysis, we examine the performance of the (suboptimal) MAP estimator in lieu of the (optimal) MMSE estimator \eqref{eq:delay-estimate:MMSE}. Using a similar set of steps to that in \eqref{eq:MAP:uniform} and \eqref{eq:AnalogPPM:Receiver_MaxCorr} for a Gaussian source, we arrive at 
\begin{subequations} 
\label{eq:AnalogPPM:MAP:Gaussian}
\begin{align}
    \hx^\MAP &= \argmax_{a\in\reals} \left\{\lambda\left(a\right)\right\},
\label{eq:AnalogPPM:Receiver_MaxCorr_Gaussian}
\end{align}
where 
\begin{align} 
    & \lambda(a) \triangleq R_{r, \phi}(a \Delta) - \frac{N}{4\sqrt{E}} a^2 ,
\label{eq:lambda:Gaussian}
\end{align}
\end{subequations}
and $R_{r, \phi}$ was defined in \eqref{eq:Rr_phi}.

\begin{remark}
    Since a Gaussian source has infinite support, the required overall transmission time $T$ is infinite. Of course this is not possible in practice. Instead, one may limit the transmission time $T$ to a very large---yet finite---value. This will incur a loss compared to the the bound that will be stated next; this loss can be made arbitrarily small by taking $T$ to be large enough.
\end{remark}

\begin{prop}
\label{prop:UpperBound_GaussianPrior}
    The distortion of the MAP estimator \eqref{eq:AnalogPPM:Receiver_MaxCorr_Gaussian} of a standard Gaussian scalar source  transmitted using analog PPM with a rectangular pulse is bounded from above as in \eqref{eq:UpperBounf_UniformPrior_Explicit}, 
    with\footnote{The notation $P_L D_L$ is used for consistency with \propref{prop:UpperBound_UniformPrior}.}
    \begin{align}
        P_L D_L &\triangleq 2\beta\sqrt{\ENR}\e^{-\frac{\ENR}{2}}\bigg(1 + 3\sqrt{\frac{2\pi}{\ENR}} + \frac{12\e^{-1}}{\beta\sqrt{\ENR}} 
      \col{}{\\* &\qquad} + \frac{8\e^{-1}}{\sqrt{8\pi}\beta} + \sqrt{\frac{8}{\pi\ENR}} + \frac{12^{\frac{3}{2}}\e^{-\frac{3}{2}}}{\beta\sqrt{32\pi\ENR}}\bigg)
     \\* &\qquad + \beta\sqrt{8\pi}\e^{-\ENR}\left(1 + \frac{4\e^{-1}}{\beta\sqrt{2\pi}}\right),
     \\ D_S &\triangleq \frac{\frac{13}{8} + \sqrt{\frac{2}{\beta}}\cdot\left(\sqrt{2 \beta \ENR} - 1 \right) \cdot \e^{-\ENR\left(1 - \frac{1}{\sqrt{2 \beta \ENR}}\right)^2} }{\left(\sqrt{\beta \ENR} - \frac{1}{\sqrt{2}} \right)^4} 
    \col{}{\\* &\qquad} + \frac{\e^{-\beta\ENR}}{\beta^2} ,
    \end{align}
    assuming $\beta \ENR > 1/2$.
    In particular, in the limit of large $\ENR$, and $\beta$ that increases monotonically with $\ENR$, 
    \begin{align}
     \label{eq:UpperBounf_GaussianPrior_Explicit_HighENR}
         D \leq \left(\tD_S + \tD_L \right)\{1 + o(1)\}
    \end{align}
    where  
    \begin{align}
         \tD_S &\triangleq\frac{13/8}{\left(\beta\ENR\right)^2},\\
         \tD_L &\triangleq 2\beta\sqrt{\ENR}\cdot\e^{-\frac{\ENR}{2}} ,
    \end{align} 
    and $o(1) \to 0$ in the limit of $\ENR \to \infty$.
\end{prop}

\begin{IEEEproof}
    The proof follows the same lines as that of \propref{prop:UpperBound_UniformPrior}. 
    
    Following Ziv and Zakai~\cite{ZivZakai:ThresholdPPM_Rect}, 
    we divide the real line into 
    consecutive intervals of length $\frac{1}{\beta}$: 
    \begin{align} 
        B_i &\triangleq \left\{ \frac{i-1}{\beta} < \eps \leq \frac{i}{\beta} \right\}, & i \in \ints. 
    \end{align}
    and define 
    \begin{align}
        A_i & \triangleq \left\{ \lambda(x) \leq \max_{a:\ \frac{i-1}{\beta} < x - a \leq \frac{i}{\beta}} \lambda(a) \right\}, & i \in \ints ,
    \label{eq:def:Ai}
    \end{align}
    where $\lambda$ was defined in \eqref{eq:lambda:Gaussian}.
    
    By using the law of total probability, we have 
    \begin{subequations}
    \label{eq:TotalProb_Error} 
    \noeqref{eq:TotalProb_Error:step1,eq:TotalProb_Error:step2,eq:TotalProb_Error:step3,eq:TotalProb_Error:step4,eq:TotalProb_Error:step5}
    \begin{align}
        \E{\eps^2} &= \sum_{i = \infty}^{\infty}\CE{\eps^2}{\eps \in B_i} \PR{\eps \in B_i}
    \label{eq:TotalProb_Error:step1}
     \\ &\leq \CE{\eps^2}{\abs{\eps} \leq \frac{1}{\beta}} + 2\sum_{i = 2}^\infty \CE{\eps^2}{\eps \in B_i}\PR{\eps \in B_i} \qquad
    \label{eq:TotalProb_Error:step2}
     \\ &\leq \CE{\eps^2}{\abs{\eps} \leq \frac{1}{\beta}} + 2\sum_{i = 2}^{\infty} \left(\frac{i}{\beta}\right)^2\E{\CPR{\eps \in B_i}{x}}
     \label{eq:TotalProb_Error:step3}
     \\ &\leq \CE{\eps^2}{\abs{\eps} \leq \frac{1}{\beta}} + 2\sum_{i = 2}^{\infty} \left(\frac{i}{\beta}\right)^2\E{\CPR{A_i}{x}} 
    \label{eq:TotalProb_Error:step4}
  \\ &\leq D_S + P_L D_L ,
    \label{eq:TotalProb_Error:step5}
    \end{align}
    \end{subequations}
    where 
    \eqref{eq:TotalProb_Error:step4} holds since $\CPR{A_i}{x} \geq \CPR{B_i}{x}$ for all $x$,
    and the two terms in \eqref{eq:TotalProb_Error:step4} are proved to be bounded from above by the two terms in \eqref{eq:TotalProb_Error:step5} in Apps.~\ref{app:SmallError_Gaussian} and \ref{app:FinalMSE_Gaussian}.
\end{IEEEproof}

\begin{thm}
\label{thm:analog-PPM:knownENR:Gaussian:achievable}
    The achievable distortion of a standard Gaussian scalar source transmitted over an energy-limited channel with a known ENR is bounded from above as 
    \begin{align}
    \label{eq:KnownENR:RectPulse_UpperBoundOptim_Final:GaussianSource}
        D &\leq 3\cdot\left(\frac{13}{8}\right)^{\frac{1}{3}}\, \e^{-\frac{\ENR}{3}}\cdot \left(\ENR\right)^{-\frac{1}{3}} \cdot \left\{1 + o(1)\right\}
        ,
    \end{align}
    where 
    $o(1) \to 0$ as $\ENR \to \infty$.
\end{thm} 

\begin{IEEEproof}
    Setting $\beta = \left(\frac{13}{8}\right)^{\frac{1}{3}}\left(\ENR\right)^{-\frac{5}{6}} \e^{\frac{\ENR}{6}}$
    in \eqref{eq:UpperBounf_GaussianPrior_Explicit_HighENR} of \propref{prop:UpperBound_GaussianPrior} yields \eqref{eq:KnownENR:RectPulse_UpperBoundOptim_Final:GaussianSource}.
\end{IEEEproof}


\section{Simulations}
\label{s:numeric}

We now present the (non-asymptotic) performance of the analog PPM scheme and compare it to that of the derived upper bound for this scheme in this work.

To that end, we simulated the analog PPM scheme for different ENR values, and optimized its performance numerically over $\beta$ for each ENR value using a Monte Carlo simulation with $10^4$ runs; for simulation purposes, the time grid was quantized uniformly with a resolution (distance between quantization points) of $\frac{1}{250\beta}$. We used $\Delta = 1$ in  \eqref{eq:AnalogPPM:PulseShaping_RectPulse}.
For the simulation for a Gaussian source, we used $T = 6.35$. None of the analog PPM pulses that resulted from the simulated random source samples exceeded this value.

\begin{figure}[t]
		\centering
	    \includegraphics[width=\columnwidth]{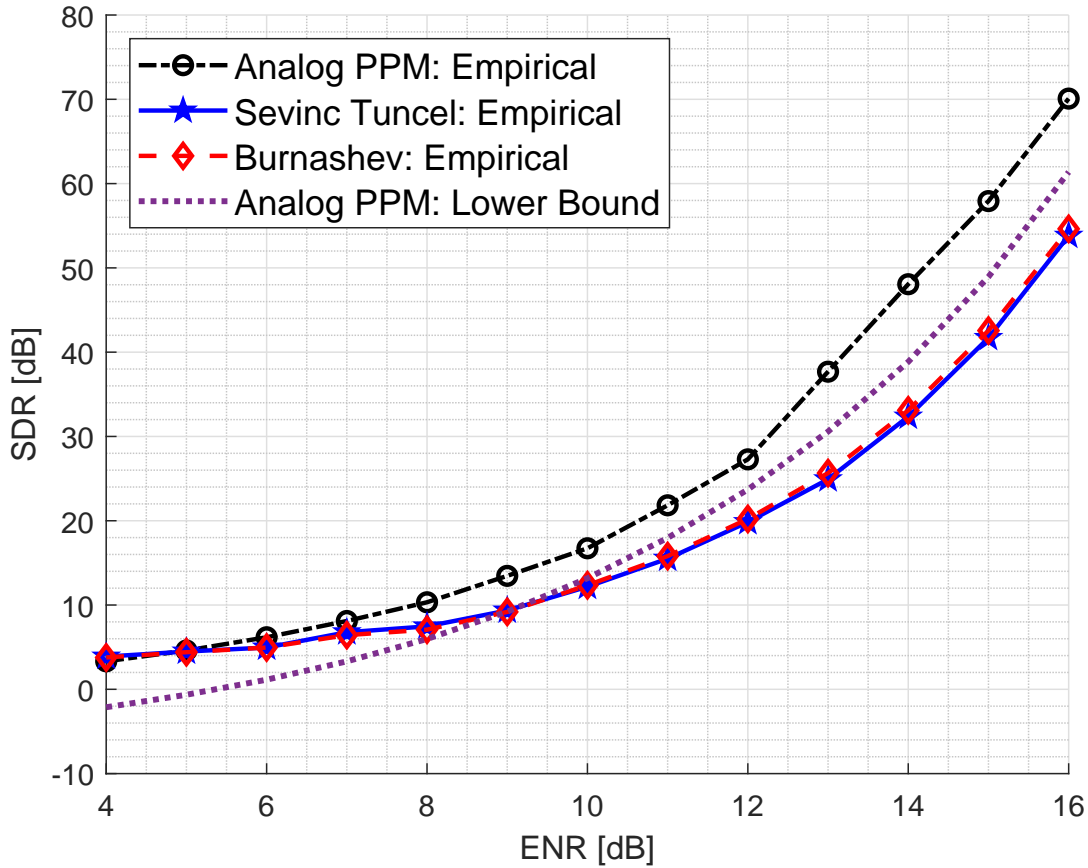}
    \caption{The optimized empirical SDR and the lower bound on the SDR of \propref{prop:UpperBound_UniformPrior}
	of analog PPM for a uniform source.}
	\label{fig:AnalogPPM_Performance}
\end{figure}

\begin{figure}[t]
		\centering
	    \includegraphics[width=\columnwidth]{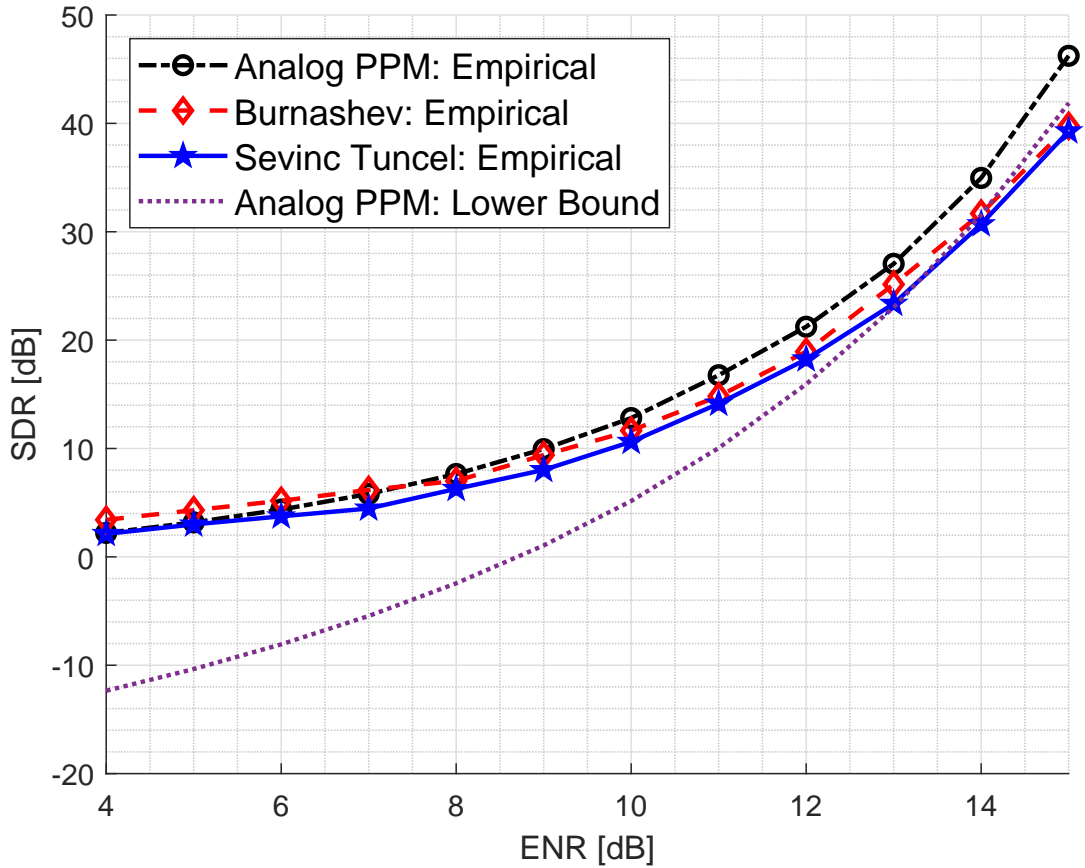}
    \caption{The optimized empirical SDR and the lower bound on the SDR of \propref{prop:UpperBound_GaussianPrior}
	of analog PPM for a standard Gaussian source 
	.}
	\label{fig:AnalogPPM_Performance_Gaussian}
\end{figure}

The numerically-optimized performance for uniform and Gaussian distributions along with the optimized bounds of \propref{prop:UpperBound_UniformPrior} and \propref{prop:UpperBound_GaussianPrior} over $\beta$ are depicted in \figref{fig:AnalogPPM_Performance} and \figref{fig:AnalogPPM_Performance_Gaussian}, respectively. 
Evidently, there is a slack in the bound which is discussed in \secref{s:Summary}.

We further compare the performance to that of the separation-based uniform-quantizer scheme of Burnashev \cite{BurnashevInfiniteBandwidthExponent1}, and to the optimized non-uniform scheme (materialized via companding) of Sevin\c{c} and Tuncel~\cite{TuncelInfinitedBW_SeparationCompanding:Journal}.
For the simulation of Burnashev's scheme for a Gaussian source, for each choice of $\beta$, we numerically optimized the overload factor and the distance between the quantized values. 
For the simulation of the scheme of Sevin\c{c} and Tuncel for a uniform source, we used the parameters from \cite{TuncelInfinitedBW_SeparationCompanding:Conf}, while for the simulation for a Gaussian source we used the $\lambda$-optimal approach of \cite{TuncelInfinitedBW_SeparationCompanding:Journal}. We used the same parameter $\beta$ for both schemes, and determined its value according to \cite[Eq.~(12)]{TuncelInfinitedBW_SeparationCompanding:Journal}.

The simulation illustrates that the proposed analog PPM scheme outperforms the separation-based schemes (with or without companding) except maybe for low ENR values. In fact, even the devised lower bound on the performance of the analog PPM outperforms the empirical performance of the separation-based schemes. Surprisingly, Burnashev's scheme seems to outperform the scheme of Sevin\c{c} and Tuncel for most of the simulated ENR values in Figs.~\ref{fig:AnalogPPM_Performance} and \ref{fig:AnalogPPM_Performance_Gaussian}, including in the (relatively) high ENR regime. This point is discussed further in \secref{s:Summary}.

Moreover, as suggested by Props.~\ref{prop:UpperBound_UniformPrior} and \ref{prop:UpperBound_GaussianPrior}, for a fixed $\beta$ value that is optimal say for some design ENR denoted by $\ENR_0$, the distortion of the analog PPM scheme decays quadratically with the ENR for $\ENR \geq \ENR_0$. This is illustrated in \figref{fig:AnalogPPM_FixedBeta} for a standard Gaussian source, where the empirical performance of the analog PPM is compared to the upper (achievability) bound of \propref{prop:UpperBound_GaussianPrior} along with the asymptotic bound of \eqref{eq:UpperBounf_UniformPrior_Explicit2}.

for $\beta = 3.68$ which is (approximately) the optimal $\beta$ value according to \thmref{thm:analog-PPM:knownENR:Gaussian:achievable} for $\ENR = 13.5 \dB$.

and its implications are discussed in \secref{s:Summary}.

\section{Discussion and Future Research}
\label{s:Summary}

In this work, we studied the problem of JSCC for a single source sample over an energy-limited AWGN channel with unlimited bandwidth and transmission time. 
We showed that analog PPM that maps directly the source sample to a time shift of the transmission pulse, 
offers improvement over separation-based schemes due to the low-delay nature of the problem on the transmitter side \cite{GastparToCodeOrNot,KochmanWornell:excess-distortion:one-shot:Allerton2015,JSCC4Control:AC2019}.

We note that, although we assumed that both the bandwidth and the time are unlimited, the scheme and analysis presented in this work carry over to the setting where one of the two is bounded as long as the other one is unlimited.
In fact, the parameter $T$ was inconsequential in the analysis and performance of the scheme, as the presented scheme may be adjusted for the setting of limited bandwidth and unlimited transmission time to achieve the same performance guarantees by proper scaling.

For a uniform source, Burnashev's impossibility bound \eqref{eq:D:Burnashev} holds with $K_1 = K_2$. Comparing it to the derived achievability bound of \thmref{thm:analog-PPM:knownENR:uniform:achievable} in this work, 
suggests that there is still a gap between the two bounds, since $K_1 < K_2$ in the latter.\footnote{Burnashev considered the problem of estimating an unknown parameter $x \in [0, 1]$. This bounds lends itself to the case of a random uniform source with the constants $K_1$ and $K_2$ remaining equal.}

In fact, Ibragimov and Khas'minskii~\cite{Ibragimov_Khasminskii:TOA_bound:PPI1975} (see also the nice summary in \cite{Zehavi:TOAbound}) derived the asymptotic performance of the MMSE estimator \eqref{eq:delay-estimate:MMSE} and that of the MAP/ML estimator \eqref{eq:MAP:uniform}, and showed that the latter is strictly suboptimal. Thus, careful analysis of the MMSE estimator in lieu of the MAP one, that was analyzed in this work, should yield better performance. 

As was noted in \secref{s:numeric}, the separation-based uniform-quantization scheme of Burnashev seems to outperform the non-uniform--quantization scheme of Sevin\c{c} and Tuncel. Since there is no apparent reason for uniform quantization to be optimal, optimization of the separation-based scheme (and quantizer therein) deserves further scrutiny. We note that we have used ML decoding in our simulation, which is clearly suboptimal. Moreover, to the best of our knowledge no (non-trivial) lower bound on the achievable distortion of separation-based schemes; it would be interesting to derive such a bound and determine whether indeed no polynomial decay (beyond the already established optimal exponential decay) may be attained using separation-based schemes.

Another interesting future research direction is deriving the optimal coefficients in Burnashev's impossibility bound.

In this work we concentrated on the setting of transmitting a single source sample over the channel. 
An interesting research direction would be to construct a non-trivial extension of the proposed scheme to vector sources.

When the noise level is not known at the transmitter (or when broadcasted to multiple parties with different ENRs) and for large memoryless Gaussian source blocklengths [for which separation becomes optimal, \ie, achieves \eqref{eq:SeparationBound}], universal schemes that attain a desired distortion profile have been devised in \cite{KokenTuncel,baniasadi2020minimum}. All of these schemes employ digital and analog linear layers for transmission:
In the infinite blocklength regime, digital layers that attain \eqref{eq:SeparationBound} may be devised for different target ENRs; however, the performance of such separation-based digital layers saturates when the true ENR is higher. Linear layers do not saturate with the ENR, but achieve only linear (albeit universal) improvement with ENR.

\begin{figure}[t]
		\centering
	    \includegraphics[width=\columnwidth]{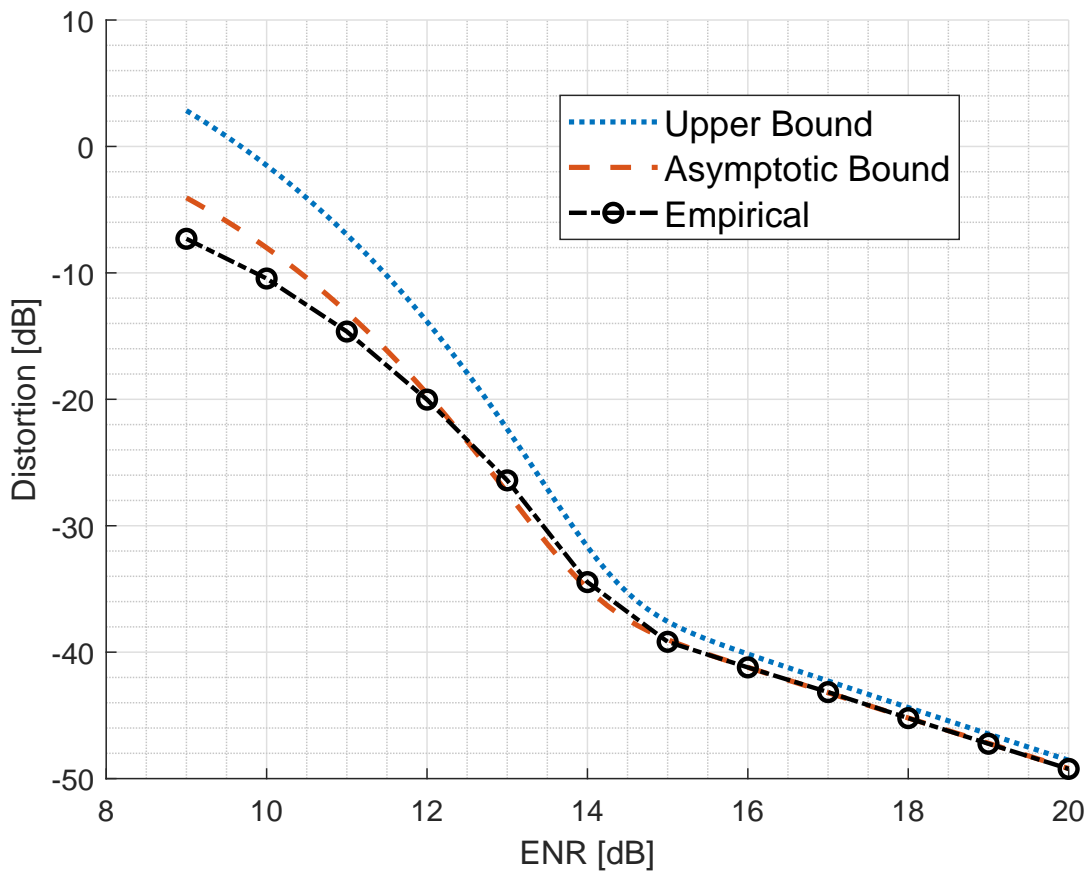}
    \caption{The empirical distortion, the upper bound of \propref{prop:UpperBound_GaussianPrior} and the asymptotic bound of \eqref{eq:UpperBounf_UniformPrior_Explicit2} of analog PPM for a standard Gaussian source. $\beta = 3.68$ was used in the simulations.}
	\label{fig:AnalogPPM_FixedBeta}
\end{figure}

Since analog PPM layers improve \textit{quadratically} with the ENR above their design ENR (see \secref{s:numeric} and \figref{fig:AnalogPPM_FixedBeta}), replacing the linear analog layers in the aforementioned universal schemes (except maybe the first linear layer) will results an improvement in the overall performance of these schemes.
This direction is pursued in a companion paper \cite{EnergyLimitedJSCC:Universal:Lev_Khina:Full}, along with an additional improvement that is offered by replacing the digital layers with modulo-lattice modulation \cite{Reznic}.

\section*{Acknowledgment}

The authors thank an anonymous reviewer of 
\cite{EnergyLimitedJSCC:Lev_Khina:ITW2021}
for helpful comments.


\appendices

\section{Upper Bound on the Small-Error Distortion in \propref{prop:UpperBound_GaussianPrior}}
\label{app:SmallError_Gaussian}

In this appendix, we bound the small-error distortion $\CE{\eps^2}{|\eps| \leq 1/\beta}$ 
for a Gaussian source. 
Our analysis builds on that of Zehavi \cite{Zehavi:TOAbound}. 
We first analyze the distortion given $x$. 

When the estimate $\hx$ deviates from $x$ by no more than $1/\beta$, 
the MAP decoder of $x$ given $r$ may be expressed 
as 
\begin{subequations}
\label{eq:SmallError_EstMetric}
\begin{align}
\hx &= \argmax_{\hx:\ |x-\hx| \leq \frac{1}{\beta}} \Bigg\{ \sqrt{E} \left( 1 - \beta|x-\hx| \right) + \sqrt{\frac{\beta}{\Delta}} \int_{\hx \Delta - \frac{\Delta}{2\beta}}^{\hx \Delta + \frac{\Delta}{2\beta}} n(t) dt 
 \col{}{\\ & \qquad\qquad\qquad\qquad\qquad\qquad\qquad} - \frac{N}{4\sqrt{E}}\hx^2 \Bigg\}
 \label{eq:SmallError_EstMetric_1}
 \\ &= \underset{\hx:\ \abs{\hx - x}\leq \frac{1}{\beta}}{\argmax} \Bigg\{-\sqrt{2\beta \ENR} \abs{\hx - x} - \frac{\hx^2}{\sqrt{8\beta\ENR}}
\col{}{\\* &\qquad\qquad\qquad} + \int_{ \hx - \frac{1}{2\beta}}^{\hx + \frac{1}{2\beta}} \tn(t)dt\Bigg\}
 \label{eq:SmallError_EstMetric_3}
 \\ &= \underset{\hx:\ \abs{\hx - x}\leq \frac{1}{\beta}}{\argmax} \bigg\{-\sqrt{2\beta \ENR}\abs{\hx - x} + \frac{x^2 - \hx^2}{\sqrt{8\beta\ENR}} 
\col{}{\\* &\qquad} + B\left(\hx + \frac{1}{\beta}\right) - B(\hx) + B(x) - B\left(x + \frac{1}{\beta}\right) \bigg\} \col{\quad\ \ }{}
\label{eq:SmallError_EstMetric_5}
\end{align}
\end{subequations}
where \eqref{eq:SmallError_EstMetric_1} 
follows from \eqref{eq:AnalogPPM:MAP:Gaussian} and \eqref{eq:correlations};
in \eqref{eq:SmallError_EstMetric_3} and in \eqref{eq:SmallError_EstMetric_5}, 
we define the white Gaussian noise $\tn(t) \triangleq \sqrt{\frac{2}{N}}n(t)$ with unit level and normalize the integration interval by factor $\Delta$, we define the two-sided Wiener process $B(x) = \int_0^x \tn(t) dt$, 
that is, $B(x) = W_1(x)$ for $x \geq 0$ and $B(x) = W_2(-x)$ for $x < 0$, such that $W_1$ and $W_2$ are independent standard Wiener processes.

We next substitute the upper bound
\begin{align}
    \label{eq:ZehaviExtension_InEqPrior}
    \frac{x^2 - \hx^2}{\sqrt{8\beta\ENR}} \leq \frac{\abs{x\left(\hx - x\right)}}{\sqrt{2\beta\ENR}}
\end{align}
in \eqref{eq:SmallError_EstMetric_5}.
One may verify that the difference function 
\begin{align}
    \abs{x\left(\hx - x\right)} - \frac{x^2 - \hx^2}{2} 
\end{align}
is monotonically increasing in the absolute error $\abs{\hx - x}$ for a given $x$.
Consequently, the probability that the $\argmax$ in \eqref{eq:SmallError_EstMetric_5} will produce larger absolute errors is higher 
after this change, meaning that the mean quadratic error (i.e., the distortion) will increase as well. 
%

Therefore, the virtual decoder 
\begin{align}
\label{eq:ZehaviExtension_MapDecoder_Maximum_UpperBound}
    \underset{\hx:\abs{\hx - x}\leq \frac{1}{\beta}}{\argmax} \bigg\{&-\sqrt{2\beta \ENR}\left(1 - \frac{\abs{x}}{2\beta\ENR}\right)\abs{\hx - x} +
\col{}{\\* &} + B\left(\hx + \frac{1}{\beta}\right) - B(\hx) + B(x) - B\left(x + \frac{1}{\beta}\right)\bigg\}
\end{align}
incurs a larger distortion than the MAP one \eqref{eq:SmallError_EstMetric}, 
and is equivalent to the decoder in \cite[Eq.~(10)]{Zehavi:TOAbound} with 
\begin{align}
    C = \sqrt{2\beta\ENR}\left(1 - \frac{\abs{x}}{2\beta\ENR}\right).
\end{align}
Thus, using the analysis of \cite{Zehavi:TOAbound} gives rise to the following upper bound on the conditional distortion of the MAP decoder given $x$:
\begin{align}
\label{eq:def:Ds()}
\begin{aligned} 
    \col{}{&}\CE{\eps^2}{x, |\eps| \leq \frac{1}{\beta}} \col{&}{}\leq D_{S}(x) 
 \\ &\col{}{\quad} \triangleq \frac{\frac{13}{8} + \sqrt{\frac{2}{\beta}} \cdot \left(\sqrt{2\beta\ENR} - \frac{\abs{x}}{\sqrt{2\beta\ENR}}\right)\cdot e^{-\ENR\left(1 - \frac{\abs{x}}{2\beta\ENR}\right)^2}}{\left( \sqrt{\beta \ENR} - \frac{\abs{x}}{2\sqrt{\beta\ENR}} \right)^4}.
\end{aligned} 
\end{align}

Consider now the probability of a small error event given~$x$:
\begin{align}
    &\CPR{|\eps| \leq \frac{1}{\beta}}{x = a}
 \\ &= \Pr\Bigg( \max_{\eps:\ \abs{\eps}\leq \frac{1}{\beta}} \left\{ z(a - \eps) - \frac{(a - \eps)^2}{\sqrt{8\beta\ENR}} + \sqrt{\frac{2\ENR}{\beta}} \left(1 - \beta\abs{\eps}\right) \right\}
\col{}{\nonumber
\\* &\qquad\qquad} \geq \max_{\eps:\ \abs{\eps} > \frac{1}{\beta}} \left\{ z(a - \eps) - \frac{(a - \eps)^2}{\sqrt{8\beta\ENR}} \right\} \Bigg) \col{\qquad}{}
\label{eq:SmallError_ConditionedProb}
\end{align}
where we defined $z(\tau) \triangleq \int_{ \tau - \frac{1}{2\beta}}^{\tau + \frac{1}{2\beta}} \tn(t)dt$.
Since $\tn$ is stationary, so is $z$. In particular, $\{z(a-\eps)| |\eps| \leq 1/\beta\}$ and $\{z(a-\eps)| |\eps| > 1/\beta\}$ are statistically identical for any $a \in \reals$.
This suggests, in turn, that $\CPR{|\eps| \leq \frac{1}{\beta}}{x = a}$ is monotonically decreasing in~$|a|$.

We are now ready to bound $\CPR{x > \sqrt{2 \beta \ENR}}{|\eps| \leq \frac{1}{\beta}}$:
\begin{subequations}
\label{eq:large_x_given_small_eps}
\noeqref{eq:large_x_given_small_eps:bayes}
\begin{align}
    \col{}{&} \CPR{x > \sqrt{2 \beta \ENR}}{|\eps| \leq \frac{1}{\beta}} 
 \col{&}{\\ &\quad} = \frac{\PR{x > \sqrt{2 \beta \ENR}}\CPR{|\eps| \leq \frac{1}{\beta}}{x > \sqrt{2 \beta \ENR}}}{\PR{\abs{\eps} \leq \frac{1}{\beta}}} \quad
\label{eq:large_x_given_small_eps:bayes}
 \\ &\col{}{\quad} \leq \PR{x > \sqrt{2\beta\ENR}}
\label{eq:large_x_given_small_eps:monotonicity}
 \\ &\col{}{\quad} < \frac{1}{2}\e^{-\beta\ENR}
\label{eq:large_x_given_small_eps:Qfunc}
\end{align}
\end{subequations}
where 
\eqref{eq:large_x_given_small_eps:monotonicity} follows from the monotonicity in $a$ of \eqref{eq:SmallError_ConditionedProb} and the law of total probability, 
and \eqref{eq:large_x_given_small_eps:Qfunc} holds since $x$ is standard Gaussian
[see, \eg, \eqref{eq:Qfunc:bounds} in \appref{app:FinalMSE_Gaussian}].

We are finally ready to bound the small-error distortion.
\begin{subequations}
\label{eq:Averaging_SmallErrors}
\noeqref{eq:Averaging_SmallErrors:Ds()}
\begin{align}
    \col{}{&}\CE{\eps^2}{|\eps| \leq \frac{1}{\beta}} 
 \col{}{\\} &= \CE{\eps^2}{|\eps| \leq \frac{1}{\beta}, |x| \leq \sqrt{2 \beta \ENR}} \CPR{|x| \leq \sqrt{2 \beta \ENR}}{|\eps| \leq \frac{1}{\beta}}
\nonumber
\\* &+ \CE{\eps^2}{|\eps| \leq \frac{1}{\beta}, |x| > \sqrt{2 \beta \ENR}} \CPR{|x| > \sqrt{2 \beta \ENR}}{|\eps| \leq \frac{1}{\beta}} 
\label{eq:Averaging_SmallErrors:integral}
 \\ &\leq D_S \left( \sqrt{2\beta\ENR} \right) + \frac{2}{\beta^2}\CPR{x > \sqrt{2 \beta \ENR}}{|\eps| \leq \frac{1}{\beta}}   
\label{eq:Averaging_SmallErrors:Ds()}
 \\ &\leq D_S \left( \sqrt{2\beta\ENR} \right) + \frac{\e^{-\beta\ENR}}{\beta^2}
\label{eq:Averaging_SmallErrors:P(x>...)}
 \\ &= D_S,
\label{eq:Averaging_SmallErrors:Ds}
\end{align}
\end{subequations}
where 
\eqref{eq:Averaging_SmallErrors:integral} follows from \eqref{eq:def:Ds()}
and the increasing monotonicity of $D_S(x)$ in $|x|$ in the region $\left\{ x: |x| \leq \sqrt{2 \beta \ENR} \right\}$ and from symmetry, 
\eqref{eq:Averaging_SmallErrors:P(x>...)} follows from \eqref{eq:large_x_given_small_eps},
and \eqref{eq:Averaging_SmallErrors:Ds} follows from the definition of $D_S$ in \propref{prop:UpperBound_GaussianPrior}.


\section{Upper Bound on the Large-Error Distortion in \propref{prop:UpperBound_GaussianPrior}}
\label{app:FinalMSE_Gaussian}


In this appendix, we will make frequent use of the following standard upper and lower bounds \cite{Q-function:approx:com1979} on 
\begin{align}
    \frac{a}{1 + a^2}\cdot \frac{\e^{-a^2/2}}{\sqrt{2\pi}} < Q(a) < \half \e^{-a^2/2},
\label{eq:Qfunc:bounds}
\end{align}

We prove here that 
\begin{align}
\label{eq:LargeErrorBound_InitSum}
    2\sum_{i = 2}^{\infty} \left(\frac{i}{\beta}\right)^2\E{\CPR{A_i}{x}} \leq P_L D_L .
\end{align}

To that end, we first bound from above the probability of $A_i$ given $x$ for $i \geq 2$. 
In the following set of inequalities, we do not write explicitly the conditioning on $x$, to simplify notation.
\vspace{-.5\baselineskip}
\begin{subequations}
\label{eq:MaximumOutOfInterval_Bound}
\noeqref{eq:MaximumOutOfInterval_Bound_step4}
\begin{align}
    \col{}{&}\PR{A_i}
    \col{&}{} = \PR{\lambda(x) \leq \max_{a:\ \frac{i-1}{\beta} < x - a \leq \frac{i}{\beta}} \lambda(a)}
\label{eq:MaximumOutOfInterval_Bound_step1}
 \\ &\leq \Pr \Bigg(\sqrt{E} + \sqrt{\frac{\beta}{\Delta}} \int_{x\Delta - \frac{\Delta}{2\beta}}^{x\Delta + \frac{\Delta}{2\beta}} n(t) dt - \frac{N}{4\sqrt{E}}x^2
\nonumber
 \\* &\qquad \leq \max_{a:\ \frac{i-1}{\beta} < x - a \leq \frac{i}{\beta}} \sqrt{\frac{\beta}{\Delta}} \int_{a\Delta - \frac{\Delta}{2\beta}}^{a\Delta + \frac{\Delta}{2\beta}} n(t) dt - \frac{N}{4\sqrt{E}} a^2 \Bigg) \quad\ \ 
\label{eq:MaximumOutOfInterval_Bound_step2}
\\  &\leq \Pr\left( \eta \leq \max_{a:\ \frac{i-1}{\beta} < x - a \leq \frac{i}{\beta}} \sqrt{\frac{\Delta}{\beta}} \int_{-\frac{1}{2}}^{\frac{1}{2}} \sqrt{\frac{2}{N}}n\left(a\Delta + \frac{\Delta}{\beta}\tau\right) d\tau \right)\ 
\label{eq:MaximumOutOfInterval_Bound_step3}
 \\ &\triangleq \Pr\left( \eta \leq \max_{0 \leq t \leq 1} w(t) \right)
 \label{eq:MaximumOutOfInterval_Bound_step4}
 \\ &= 1 -  \E{\CPR{\max_{0 < t \leq 1} w(t) < \eta}{\eta}}
\label{eq:MaximumOutOfInterval_Bound_step5}
 \\ &= 1 -  \E{\left[1 - Q(\eta)\right]^2 - \frac{\eta \e^{-\frac{\eta^2}{2}}}{\sqrt{2\pi}}\left[1 - Q(\eta)\right] - \frac{\e^{-\eta^2}}{2\pi}}
\label{eq:MaximumOutOfInterval_Bound_step6}
 \\ &< \E{2Q(\eta) + \frac{\eta \e^{-\frac{\eta^2}{2}}}{\sqrt{2\pi}}\left[1 - Q(\eta)\right] + \frac{\e^{-\eta^2}}{2\pi}}
\label{eq:MaximumOutOfInterval_Bound_step7}
\\ &\leq \frac{\sqrt{3}}{4\pi}\e^{-\frac{q^2(x)}{3}} + 2Q(q(x)) 
    + \frac{q(x)\e^{-\frac{q^2(x)}{4}}}{4\sqrt{\pi}}\left(1 - Q\left(\frac{q(x)}{\sqrt{6}}\right)\right)
\col{}{\\& \qquad\quad} + \frac{\e^{-\frac{q^2(x)}{4}}}{2\sqrt{2}}\left(1 - Q\left(\frac{q(x)}{\sqrt{2}}\right)\right) \quad\ \ 
\label{eq:MaximumOutOfInterval_Bound_step8}
\\ &\leq \frac{\sqrt{3}}{4\pi}\e^{-\frac{\ell^2(x)}{3}} + 2Q(\ell(x)) 
    + \frac{\ell(x)\e^{-\frac{\ell^2(x)}{4}}}{4\sqrt{\pi}}\left(1 - Q\left(\frac{\ell(x)}{\sqrt{6}}\right)\right)
\col{}{\\* &\qquad\quad} + \frac{\e^{-\frac{\ell^2(x)}{4}}}{2\sqrt{2}}\left(1 - Q\left(\frac{\ell(x)}{\sqrt{2}}\right)\right)    
\label{eq:MaximumOutOfInterval_Bound_step9}
\\ &\leq          
\begin{cases}
    1, & \ell(x) < 0
 \\ \left(\frac{1}{2\sqrt{2}} + \frac{\ell(x)}{4\sqrt{\pi}}\right)\e^{-\frac{\ell^2(x)}{4}} 
  + \e^{-\frac{\ell^2(x)}{2}} + \frac{\sqrt{3}}{4\pi}\e^{-\frac{\ell^2(x)}{3}}, & \ell(x) \geq 0
\end{cases}\ \ 
\label{eq:MaximumOutOfInterval_Bound_step10}
\end{align}  
\end{subequations}
where 
\eqref{eq:MaximumOutOfInterval_Bound_step1} and \eqref{eq:MaximumOutOfInterval_Bound_step2} hold by the definitions of $A_i$ \eqref{eq:def:Ai} and $\lambda$ \eqref{eq:lambda:Gaussian}, respectively, and since 
$R_{r,\phi}(a \Delta) = 0$ \eqref{eq:correlations} for $i \geq 2$;
\eqref{eq:MaximumOutOfInterval_Bound_step3} holds by defining 
\begin{align}
    \eta &\triangleq \sqrt{\frac{\beta}{\Delta}} \int_{x\Delta - \frac{\Delta}{2\beta}}^{x\Delta + \frac{\Delta}{2\beta}} \sqrt{\frac{2}{N}} n(t) dt 
\col{}{\\* &\quad} + \frac{1}{\sqrt{8\ENR}} \left[ \min\left\{\left(\frac{i}{\beta} + x\right)^2,\left(\frac{i - 1}{\beta} + x\right)^2\right\} - x^2\right],
\end{align}
and by taking the minimum value of $a^2$ inside the interval $\left\{ a:\ \frac{i-1}{\beta} < x-a \leq \frac{i}{\beta} \right\}$;
\eqref{eq:MaximumOutOfInterval_Bound_step4} holds by defining the process
\begin{align}
    w(a) \triangleq \sqrt{\frac{\Delta}{\beta}} \int_{-\frac{1}{2}}^{\frac{1}{2}} \sqrt{\frac{2}{N}}n\left(a\Delta + \frac{\Delta}{\beta}\tau\right) d\tau ;
\end{align}
\eqref{eq:MaximumOutOfInterval_Bound_step6} is due to \cite{shepp1966radon} (see also \cite[Eq.~7]{ZivZakai:ThresholdPPM_Rect}), by noting that $w$ is a zero-mean Gaussian process with autocorrelation function
\begin{align}
    R_w(\tau) &=
    \begin{cases}
        1 - \abs{\tau}, & \abs{\tau} \leq 1,
     \\ 0, & \mathrm{otherwise};
    \end{cases}
\end{align}    
\eqref{eq:MaximumOutOfInterval_Bound_step7} holds by adding $\E{Q^2(\eta)} > 0$;
\eqref{eq:MaximumOutOfInterval_Bound_step8} follows from \cite[App.~A]{ZivZakai:ThresholdPPM_Rect} and
\begin{align}
\col{}{&}\frac{1}{\sqrt{2\pi}}\int_{-\infty}^{\infty}Q(x)\e^{-\frac{\left(x - a\right)^2}{2}}dx 
\col{}{\\}
    &\leq \frac{1}{\sqrt{2\pi}}\left(\int_{-\infty}^{0}\e^{-\frac{\left(x - a\right)^2}{2}}dx + \frac{\e^{-\frac{a^2}{4}}}{2}\int_{0}^{\infty}\e^{-\left(x - \frac{a}{2}\right)^2}dx\right)\\
    &= Q(a) + \frac{\e^{-\frac{a^2}{4}}}{2\sqrt{2}}\left(1 - Q\left(\frac{a}{\sqrt{2}}\right)\right) ,
\end{align}
and by defining
\begin{align}
    q(x) &\triangleq \sqrt{2\ENR}\bigg(1  - \frac{x^2}{4\ENR} 
\col{}{\\* &\quad\qquad} + \frac{1}{4\ENR}\min\left\{\left(\frac{i}{\beta} + x\right)^2,\left(\frac{i - 1}{\beta} + x\right)^2\right\}\bigg)
 \\ &= \sqrt{2\ENR}\left(1 + \frac{1}{4\ENR}\bigg(\frac{i}{\beta}\right)^2 + \frac{1}{2\ENR}\frac{i}{\beta}x
\col{}{\\* &\quad\qquad} + \frac{1}{4\ENR}\min\left\{0,\frac{1}{\beta^2} - \frac{2}{\beta}\left(x + \frac{i}{\beta}\right)\right\}\bigg);
\end{align}
\eqref{eq:MaximumOutOfInterval_Bound_step9} holds by the monotonicity of the expression in \eqref{eq:MaximumOutOfInterval_Bound_step8} with respect to $q$ for $q>0$, which can be verified by differentiating it with respect to $q$ and prove that it is negative for all $q$ 
using \eqref{eq:Qfunc:bounds} 
by defining 
\begin{align}
    &\ell(x) \triangleq \sqrt{2\ENR}\left(1 + \frac{1}{4\ENR}\left(\frac{i}{\beta}\right)^2 + \frac{1}{2\ENR}\frac{i}{\beta}x \right),
\label{eq:def:ell}
\end{align}
which by definition is always greater than or equal to $q$,
and since the probability is always bounded from above by $1$;
\eqref{eq:MaximumOutOfInterval_Bound_step10} follows from the upper bound in \eqref{eq:Qfunc:bounds},
and by dropping negative terms. 

We now bound the unconditional probability $\PR{A_i}$ using the law of total probability and \eqref{eq:MaximumOutOfInterval_Bound} [recall that, in \eqref{eq:MaximumOutOfInterval_Bound}, we bounded the conditional probability of $A_i$ given $x$]:
\begin{subequations}
\label{eq:Ai:calc}
\noeqref{eq:Ai:calc:smoothing:def,eq:Ai:calc:smoothing:explicit,eq:Ai:calc:UB1,eq:Ai:calc:UB2}
\begin{align}
    &\PR{A_i} 
    = \E{\CPR{A_i}{x=a}}
\label{eq:Ai:calc:smoothing:def}
 \\ &= \frac{1}{\sqrt{2\pi}}\int_{-\infty}^{\infty}\CPR{A_i}{x=a}\e^{-\frac{a^2}{2}}da 
\label{eq:Ai:calc:smoothing:explicit}
 \\ &\leq \frac{1}{\sqrt{2\pi}} \int_{a: \ell(a) \leq 0}\e^{-\frac{a^2}{2}} da + \frac{1}{\sqrt{2\pi}} \int_{-\infty}^{\infty}\e^{-\frac{a^2}{2}} \Bigg\{ \frac{\sqrt{3}}{4\pi}\e^{-\frac{\ell^2(a)}{3}}
\col{}{\\* &\qquad} + \left(\frac{1}{2\sqrt{2}} + \frac{\ell(a)}{4\sqrt{\pi}}\right)\e^{-\frac{\ell^2(a)}{4}} +  \e^{-\frac{\ell^2(a)}{2}} \Bigg\} da \quad\ \ 
\label{eq:Ai:calc:integral-identity}
 \\ &\leq Q\left(2\ENR\frac{\beta}{i} + \frac{i}{2\beta}\right)
\col{}{\\ &} + \frac{\frac{1}{2\sqrt{2}} + \sqrt{\frac{\ENR}{8\pi}} \left(1 + \frac{1}{4\ENR}\left(\frac{i}{\beta}\right)^2\right)}{\sqrt{1 + \frac{1}{4\ENR}\left(\frac{i}{\beta}\right)^2}}\cdot\e^{-\frac{\ENR}{2} \left(1 + \frac{1}{4\ENR}\left(\frac{i}{\beta}\right)^2\right)} 
\nonumber
\\ &\qquad + \frac{1}{\sqrt{1 + \frac{1}{2\ENR}\left(\frac{i}{\beta}\right)^2}}\cdot\e^{-\ENR\frac{\left(1 + \frac{1}{4\ENR}\left(\frac{i}{\beta}\right)^2\right)^2}{1 + \frac{1}{2\ENR}\left(\frac{i}{\beta}\right)^2}}
\col{}{\\ &\qquad} + \frac{\frac{\sqrt{3}}{4\pi}}{\sqrt{1 + \frac{1}{3\ENR}\left(\frac{i}{\beta}\right)^2}}\e^{-\frac{2\ENR}{3}\frac{\left(1 + \frac{1}{4\ENR}\left(\frac{i}{\beta}\right)^2\right)^2}{1 + \frac{1}{3\ENR}\left(\frac{i}{\beta}\right)^2}}
\label{eq:Ai:calc:plug-ell}
 \\ &\leq \left(\frac{1}{2\sqrt{2}} + \sqrt{\frac{\ENR}{8\pi} + \frac{1}{32\pi}\left(\frac{i}{\beta}\right)^2}\right)\cdot\e^{-\frac{\ENR}{2} \left(1 + \frac{1}{4\ENR}\left(\frac{i}{\beta}\right)^2\right)} 
\\ &\qquad + \e^{-\frac{\ENR}{2}\left(1 + \frac{1}{4\ENR}\left(\frac{i}{\beta}\right)^2\right)}
    + \frac{\sqrt{3}}{4\pi}\e^{-\frac{\ENR}{2}\left(1 + \frac{1}{4\ENR}\left(\frac{i}{\beta}\right)^2\right)}
\col{}{\\ &\qquad} + Q\left(2\ENR\frac{\beta}{i} + \frac{i}{2\beta}\right)
\label{eq:Ai:calc:UB1}
 \\ &\leq \left(\frac{3}{2} + \sqrt{\frac{\ENR}{8\pi} + \frac{1}{32\pi}\left(\frac{i}{\beta}\right)^2}\right)\cdot \e^{- \frac{\ENR}{2} - \frac{1}{8}\left( \frac{i}{\beta} \right)^2}
\col{}{\\ &\qquad} +  Q\left(2\ENR\frac{\beta}{i} + \frac{i}{2\beta}\right)
\label{eq:Ai:calc:UB2}
\end{align}
\end{subequations}
where \eqref{eq:Ai:calc:integral-identity} follows from \eqref{eq:MaximumOutOfInterval_Bound};
\eqref{eq:Ai:calc:plug-ell} follows from the definition of $\ell$ \eqref{eq:def:ell} 
and the following integral identity (and upper bound) which holds for any positive reals $a,b,c,k_2$ and and any real $ k_1$:
\begin{align}
    \col{}{&}\frac{1}{\sqrt{2\pi}}\int_{-\infty}^{\infty}(k_1 + k_2 u)\e^{-\frac{1}{2}\left(a\left(b + cu\right)^2 + u^2\right)}du 
 \col{}{\\} &= \frac{k_1}{\sqrt{1 + ac^2}}\e^{-\frac{ab^2}{2\left(1 + ac^2\right)}} - \frac{abck_2}{\left(1 + ac^2\right)^{\frac{3}{2}}}\e^{-\frac{ab^2}{2\left(1 + ac^2\right)}}
 \\ &\leq \frac{k_1}{\sqrt{1 + ac^2}}\e^{-\frac{ab^2}{2\left(1 + ac^2\right)}}.
\end{align}

Combining the upper bound \eqref{eq:Ai:calc} on $\PR{A_i}$ and using \eqref{eq:LargeErrorBound_InitSum}, 
we are ready to prove the desired bound:
\begin{subequations} 
\label{eq:PlDl} 
\begin{align}
    \col{}{&}2\sum_{i = 2}^{\infty} \left(\frac{i}{\beta}\right)^2 \PR{A_i}
    \col{&}{} \leq 2\sum_{i = 2}^{\infty}\left(\frac{i}{\beta}\right)^{2}
    Q\left(2\ENR\frac{\beta}{i} + \frac{i}{2\beta}\right)
\\* &\ \  + 2\sum_{i = 2}^{\infty}\left(\frac{i}{\beta}\right)^{2}\left(\frac{3}{2} + \sqrt{\frac{\ENR}{8\pi} + \frac{\left(i/\beta\right)^2}{32\pi}}\right) \e^{- \frac{\ENR}{2} - \frac{1}{8}\left( \frac{i}{\beta} \right)^2} \quad\ 
\label{eq:PlDl:final:basic}
 \\ &= 2\left(D_{L,1} + D_{L,2}\right)
\label{eq:PlDl:final:DLs}
 \\ &\leq P_L D_L,
\label{eq:PlDl:final:last}
\end{align}
\label{eq:PlDl:final}
\end{subequations}
where $D_{L,1}$ and $D_{L,2}$ in \eqref{eq:PlDl:final:DLs} denote the two sums in \eqref{eq:PlDl:final:basic}, and \eqref{eq:PlDl:final:last} is proved next.

To bound $D_{L,1}$ and $D_{L,2}$ we will make use of the following bound for $d = 0,1,2,3$.
\begin{subequations}
\label{eq:InfiniteSumBound}
\noeqref{eq:InfiniteSumBound_2,eq:InfiniteSumBound_3,eq:InfiniteSumBound_4}
\begin{align}
    \col{}{&}\sum_{i = 0}^{\infty} \left(\frac{i}{\beta}\right)^{d}\e^{-\frac{i^2}{8\beta^2}}
\col{}{\\*} &\leq \int_{0}^{\left\lceil \sqrt{4d}\beta \right\rceil}\left(\frac{u}{\beta}\right)^d\e^{-\frac{u^2}{8\beta^2}}du + \int_{\left\lfloor \sqrt{4d} \beta \right\rfloor - 1}^{\infty}\left(\frac{u}{\beta}\right)^d\e^{-\frac{u^2}{8\beta^2}}du \quad\ \ 
\label{eq:InfiniteSumBound_1}
 \\ &= \beta \int_{0}^{\infty}u^d \e^{-\frac{u^2}{8}}du 
    + \beta \int_{\frac{\left\lfloor \sqrt{4d} \beta \right\rfloor - 1}{\beta}}^{\frac{\left\lceil \sqrt{4d} \beta \right\rceil}{\beta}}
    u^d \e^{-\frac{u^2}{8}}du
\label{eq:InfiniteSumBound_2}
 \\ &\leq 2^{d + 1} \beta \int_0^{\infty} u^d \e^{-\frac{u^2}{2}} du 
    + 2 \left(\sqrt{4d}\right)^d\e^{-\frac{d}{2}} 
\label{eq:InfiniteSumBound_3}
 \col{\\*}{\\} &=
    \begin{cases}
        \sqrt{2\pi} \beta, & d = 0;
     \\ 4\beta + 4 \e^{-1/2}, & d = 1;
     \\ 4\beta\sqrt{2\pi} + 16 \e^{-1}, & d = 2;
     \\ 32\beta + 48 \sqrt{3} \e^{-3/2}, & d=3;
    \end{cases}
\label{eq:InfiniteSumBound_4}
\end{align}
\end{subequations}
where \eqref{eq:InfiniteSumBound_1} holds since the
function $x^d \e^{-\frac{x^2}{8\beta^2}}$ is monotonically increasing for $x \in \left( 0, \beta\sqrt{4d} \right)$, 
and monotonically decreasing for $x > \beta\sqrt{4d}$; we use the convention $0^0 \triangleq 0$ for the case $d = 0$ since the sum in \eqref{eq:InfiniteSumBound_1} can be bounded by the first integral in this case.

We next bound the term $D_{L,1}$ from above.
\begin{subequations}
\noeqref{eq:D1_Bound_Step0}
\label{eq:D1_Bound}
\begin{align}
    \col{}{&} D_{L,1} 
 \col{&}{} \triangleq \sum_{i = 2}^{\infty}\left(\frac{i}{\beta}\right)^{2}\left(\frac{3}{2} + \sqrt{\frac{\ENR}{8\pi} + \frac{1}{32\pi}\left(\frac{i}{\beta}\right)^2}\right) \cdot \e^{- \frac{\ENR}{2} - \frac{1}{8}\left( \frac{i}{\beta} \right)^2}
\label{eq:D1_Bound_Step0}
 \\ &\leq\e^{-\frac{\ENR}{2}}\sum_{i = 0}^{\infty}\left(\frac{i}{\beta}\right)^{2}\left(\frac{3}{2} + \sqrt{\frac{\ENR}{8\pi}}\left(1 + \frac{\frac{i}{\beta}}{2\sqrt{\ENR}}\right)\right) \cdot \e^{-\frac{1}{8}\left( \frac{i}{\beta} \right)^2} \quad\
\label{eq:D1_Bound_Step2}
 \\ &\leq 2\beta\sqrt{\ENR}\e^{-\frac{\ENR}{2}}\bigg(1 + 3\sqrt{\frac{2\pi}{\ENR}} 
    + \frac{12\e^{-1}}{\beta\sqrt{\ENR}} 
\col{}{\\* &\qquad\qquad\qquad} + \frac{8\e^{-1}}{\sqrt{8\pi}\beta} + \sqrt{\frac{8}{\pi\ENR}} 
    + \frac{12^{\frac{3}{2}}\e^{-\frac{3}{2}}}{\beta\sqrt{32\pi\ENR}}\bigg) \col{\quad}{}
\label{eq:D1_Bound_Step3}
\end{align}
\end{subequations}
where \eqref{eq:D1_Bound_Step2} follows from the inequality $\sqrt{1 + x} \leq 1 + \sqrt{x}$ that holds for $x \geq 0$ and by adding positive terms corresponding to $i = 0, 1$;
and \eqref{eq:D1_Bound_Step3} follows from \eqref{eq:InfiniteSumBound}.

Similarly, we bound $D_{L,2}$ as follows.
\begin{subequations}
\noeqref{eq:D4_Bound:def,eq:D4_Bound_Step0}
\label{eq:D4_Bound} 
\begin{align}
    D_{L,2} &\triangleq \sum_{i = 2}^{\infty}\left(\frac{i}{\beta}\right)^{2} Q\left(2\ENR\frac{\beta}{i} + \frac{i}{2\beta}\right)
\label{eq:D4_Bound:def}
 \\ &\leq \half \sum_{i = 2}^{\infty}\left(\frac{i}{\beta}\right)^{2} \e^{-\frac{1}{2}\left(2\ENR\frac{\beta}{i} + \frac{i}{2\beta}\right)^2}
\label{eq:D4_Bound_Step0}
 \\ &\leq \half \e^{-\ENR}\sum_{i = 0}^{\infty}\left(\frac{i}{\beta}\right)^{2} \e^{-\frac{1}{2}\left(\frac{i}{2\beta}\right)^2}
\label{eq:D4_Bound_Step1}
  \\ &\leq \beta\sqrt{8\pi}\e^{-\ENR}\left(1 + \frac{4\e^{-1}}{\beta\sqrt{2\pi}}\right),
\label{eq:D4_Bound_Step2}
\end{align}
\end{subequations}
where \eqref{eq:D4_Bound_Step1} follows from \eqref{eq:Qfunc:bounds}, and by dropping multiplicative terms with negative exponent; 
and \eqref{eq:D4_Bound_Step2} follows from \eqref{eq:InfiniteSumBound}. 

Combining the bounds on $D_{L,1}$ and $D_{L,2}$ 
in \eqref{eq:D1_Bound} and \eqref{eq:D4_Bound} yields \eqref{eq:PlDl:final:last} and concludes the proof.


\bibliographystyle{IEEEtran}
\bibliography{myBib}

\end{document}

%% file: defs.tex
\usepackage[tbtags]{amsmath} 
\usepackage{amssymb}  
\usepackage{verbatim} 
\usepackage{amsxtra}  
\usepackage{comment}

\usepackage{mathabx}

\usepackage{enumerate}

\usepackage{amsfonts}
\usepackage{color}

\usepackage{subcaption}

\usepackage{wasysym}

\usepackage{cite}

\usepackage{tikz}
\usetikzlibrary{shapes,arrows,decorations.markings,positioning}

\tikzstyle{plant} = [draw, fill=red!5, rectangle, 
    minimum height=3em, minimum width=6em]
\tikzstyle{block} = [draw, fill=blue!5, rectangle, 
    minimum height=3em, minimum width=6em]
\tikzstyle{sum} = [draw, fill=yellow!10, circle, node distance=1cm]
\tikzstyle{coord} = [coordinate]
\tikzstyle{gain} = [draw, fill=red!5, regular polygon, regular polygon sides=3, shape border rotate=-90]
\tikzstyle{pinstyle} = [pin edge={to-,thick,black}]

\tikzstyle{BitPipe} = [thick, decoration={markings,mark=at position
   1 with {\arrow[semithick]{open triangle 60}}},
   double distance=1.4pt, shorten >= 5.5pt,
   preaction = {decorate},
   postaction = {draw,line width=1.4pt, white,shorten >= 4.5pt}]

\usepackage{hyperref}

\usepackage{epsfig, graphicx, psfrag}

\usepackage[mathcal]{euscript}
\usepackage[cal=boondox]{mathalfa}

\providecommand{\thmref}[1]{Th.~\ref{#1}}

\providecommand{\secref}[1]{Sec.~\ref{#1}}

\providecommand{\propref}[1]{Prop.~\ref{#1}}

\providecommand{\figref}[1]{Fig.~\ref{#1}}

\providecommand{\appref}[1]{App.~\ref{#1}}

\newcommand{\ie}{i.e.}

\newcommand{\eg}{e.g.}

\newcommand{\viz}{viz.}

\newcommand{\reals}{\mathbb{R}}
\newcommand{\ints}{\mathbb{Z}}

\newcommand{\nats}{\mathbb{N}}

\DeclareMathOperator*{\argmax}{argmax}
\DeclareMathOperator*{\argmin}{argmin}

\newcommand{\ENR}{\mathrm{ENR}}

\newcommand{\SDR}{\mathrm{SDR}}

\newcommand{\e}{\mathrm{e}}

\newcommand{\Comment}[1]{}
\newcommand{\old}[1]{}
\newcommand{\rem}[1]{}

\newcommand{\eps}{{\epsilon}}

\newcommand{\MMSE}{{\mathrm{MMSE}}}

\newcommand{\hx}{{\hat{x}}}

\newcommand{\tP}{\tilde{P}}

\newcommand{\tn}{\tilde{n}}

\newcommand{\tD}{\tilde D}

\newcommand{\abs}[1]{\left| #1 \right|}
\newcommand{\Norm}[1]{\left\| #1 \right\|}

\providecommand{\e}{{\rm e}}
\providecommand{\comment}[1]{}

\providecommand{\norm}[1]{\Norm{#1}}

\newcommand{\beqn}[1]{\begin{eqnarray}\label{#1}}
\newcommand{\eeqn}{\end{eqnarray}}
\newcommand{\beq}[1]{\begin{equation}\label{#1}}
\newcommand{\eeq}{\end{equation}}

\newcommand{\MAP}{\mathrm{MAP}}

\providecommand{\half}{\frac{1}{2}}

\providecommand{\half}{\frac{1}{2}}

\makeatletter
\newcommand{\vast}{\bBigg@{4}}
\newcommand{\Vast}{\bBigg@{5}}
\makeatother

\providecommand{\dB}{\mathrm{dB}}

\providecommand{\bbE}{\mathbb{E}}

\providecommand{\E}[1]{\bbE \left[ #1 \right]}
\providecommand{\CE}[2]{\bbE \left[ #1 \middle| #2 \right]}

\renewcommand{\gamma}{\ENR}

\newcommand{\PR}[1]{\Pr\left( #1 \right)}
\newcommand{\CPR}[2]{\Pr\left( #1 \middle| #2 \right)}

%% file: known_ENR_journal.bbl
\begin{thebibliography}{10}
\providecommand{\url}[1]{#1}
\csname url@samestyle\endcsname
\providecommand{\newblock}{\relax}
\providecommand{\bibinfo}[2]{#2}
\providecommand{\BIBentrySTDinterwordspacing}{\spaceskip=0pt\relax}
\providecommand{\BIBentryALTinterwordstretchfactor}{4}
\providecommand{\BIBentryALTinterwordspacing}{\spaceskip=\fontdimen2\font plus
\BIBentryALTinterwordstretchfactor\fontdimen3\font minus
  \fontdimen4\font\relax}
\providecommand{\BIBforeignlanguage}[2]{{%
\expandafter\ifx\csname l@#1\endcsname\relax
\typeout{** WARNING: IEEEtran.bst: No hyphenation pattern has been}%
\typeout{** loaded for the language `#1'. Using the pattern for}%
\typeout{** the default language instead.}%
\else
\language=\csname l@#1\endcsname
\fi
#2}}
\providecommand{\BIBdecl}{\relax}
\BIBdecl

\bibitem{CoverBook2Edition}
T.~M. Cover and J.~A. Thomas, \emph{Elements of Information Theory, Second
  Edition}.\hskip 1em plus 0.5em minus 0.4em\relax New York: Wiley, 2006.

\bibitem{ElGamalKimBook}
A.~{El Gamal} and Y.-H. Kim, \emph{Network Information Theory}.\hskip 1em plus
  0.5em minus 0.4em\relax Cambridge University Press, 2011.

\bibitem{Shannon59:RDF}
C.~E. Shannon, ``Coding theorems for a discrete source with a fidelity
  criterion,'' in \emph{Institute of Radio Engineers, International Convention
  Record}, vol.~7, 1959, pp. 142--163.

\bibitem{Cohn_Phd}
D.~L. Cohn, ``Minimum mean-square error without coding,'' Ph.D. dissertation,
  Dept.\ EECS, Massachusetts Institute of Technology, 1970.

\bibitem{BurnashevInfiniteBandwidthExponent}
M.~Burnashev, ``Minimum attainable mean-square error in transmission of a
  parameter over a channel with white {G}aussian noise,'' \emph{\emph{(in
  Russian)} Problemy Peredachi Informatsii (Problems of Information
  Transmission)}, vol.~21, 10 1985.

\bibitem{BurnashevInfiniteBandwidthExponent1}
M.~{Burnashev}, ``A new lower bound for the a-mean error of parameter
  transmission over the white {G}aussian channel,'' \emph{IEEE Transactions on
  Information Theory}, vol.~30, no.~1, pp. 23--34, 1984.

\bibitem{SeparationInfBW_Abdallah}
F.~Abi~Abdallah and R.~Knopp, ``Source--channel coding for very-low bandwidth
  sources,'' in \emph{Proceedings of the IEEE Information Theory Workshop
  (ITW)}, Guangzhou, China, 2008, pp. 184--188.

\bibitem{TuncelInfinitedBW_SeparationCompanding:Journal}
C.~Sevin{\c{c}} and E.~Tuncel, ``On asymptotic analysis of energy--distortion
  tradeoff for low-delay transmission over {Gaussian} channels,'' \emph{IEEE
  Transactions on Communications}, vol.~69, no.~7, pp. 4448--4460, 2021.

\bibitem{WozencraftJacobsBook}
J.~M. Wozencraft and I.~M. Jacobs, \emph{Principles of Communication
  Engineering}.\hskip 1em plus 0.5em minus 0.4em\relax New York: John Wiley \&
  Sons, 1965.

\bibitem{GallagerBook1968}
R.~G. Gallager, \emph{Information Theory and Reliable Communication}.\hskip 1em
  plus 0.5em minus 0.4em\relax New York: John Wiley \& Sons, 1968.

\bibitem{ViterbiOmuraBook}
A.~J. Viterbi and J.~K. Omura, \emph{Principles of Digital Communication and
  Coding}.\hskip 1em plus 0.5em minus 0.4em\relax \!\!\!New York: McGraw-Hill,
  1979.

\bibitem{Bennett48}
W.~R. Bennett, ``Spectra of quantized signals,'' \emph{Bell System Technical
  Journal}, vol.~27, pp. 446--472, Jul. 1948.

\bibitem{PanterDite:FLC}
P.~F. Panter and W.~Dite, ``Quantization distortion in pulse-count modulation
  with nonuniform spacing of levels,'' \emph{Proc.\ IRE}, vol.~39, no.~1, pp.
  44--48, Jan. 1951.

\bibitem{GershoGrayBook}
A.~Gersho and R.~M. Gray, \emph{Vector Quantization and Signal
  Compression}.\hskip 1em plus 0.5em minus 0.4em\relax Boston: Kluwer Academic
  Pub., 1992.

\bibitem{ChazanZivZakai:ParamterEstimationBound:IT1975}
D.~Chazan, M.~Zakai, and J.~Ziv, ``Improved lower bounds on signal parameter
  estimation,'' \emph{IEEE Transactions on Information Theory}, vol.~21, no.~1,
  pp. 90--93, 1975.

\bibitem{WeissWeinstein:TOAbound}
A.~J. {Weiss}, ``Composite bound on arrival time estimation errors,''
  \emph{IEEE Transactions on Aerospace and Electronics Systems}, vol.~22,
  no.~6, pp. 751--756, 1986.

\bibitem{Zehavi:TOAbound}
E.~{Zehavi}, ``Estimation of time of arrival for rectangular pulses,''
  \emph{IEEE Transactions on Aerospace and Electronics Systems}, vol.~20,
  no.~6, pp. 742--747, 1984.

\bibitem{ZivZakai:ThresholdPPM_Rect}
M.~{Zakai} and J.~{Ziv}, ``On the threshold effect in radar range estimation
  (corresp.),'' \emph{IEEE Transactions on Information Theory}, vol.~15, no.~1,
  pp. 167--170, 1969.

\bibitem{Merhav:StatPhys_ThresholdPPM}
N.~Merhav, ``Threshold effects in parameter estimation as phase transitions in
  statistical mechanics,'' \emph{IEEE Transactions on Information Theory},
  vol.~57, no.~10, pp. 7000--7010, 2011.

\bibitem{TuncelInfinitedBW_SeparationCompanding:Conf}
C.~Sevin{\c{c}} and E.~Tuncel, ``On asymptotic analysis of energy--distortion
  tradeoff for low-delay transmission over {Gaussian} channels,'' in
  \emph{Proceedings of the IEEE International Symposium on Information Theory
  (ISIT)}, 2018, pp. 2599--2603.

\bibitem{GastparToCodeOrNot}
M.~Gastpar, B.~Rimoldi, and M.~Vetterli, ``To code, or not to code: Lossy
  source--channel communication revisited,'' \emph{IEEE Transactions on
  Information Theory}, vol.~49, no.~5, pp. 1147--1158, May 2003.

\bibitem{KochmanWornell:excess-distortion:one-shot:Allerton2015}
Y.~Kochman and G.~W. Wornell, ``Excess distortion in lossy compression: Beyond
  one-shot analysis,'' in \emph{Proceedings of the Annual Allerton Conference
  on Communication, Control, and Computing}, Monticello, IL, USA, 2015, pp.
  928--934.

\bibitem{JSCC4Control:AC2019}
\khina\CoRes, E.~{Riedel G\r{a}rding}\Student, G.~M. Pettersson\Student,
  V.~Kostina\PI, and B.~Hassibi\PI, ``Control over {Gaussian} channels with and
  without source--channel separation,'' \emph{IEEE Transactions on Automatic
  Control}, vol.~64, no.~9, pp. 3690--3705, Sep. 2019.

\bibitem{Ibragimov_Khasminskii:TOA_bound:PPI1975}
I.~A. Ibragimov and R.~Z. Khas'minskii, ``Parameter estimation for a
  discontinuous signal in white {Gaussian} noise,'' \emph{\emph{(in Russian)}
  Problemy Peredachi Informatsii (Problems of Information Transmission)},
  vol.~11, no.~3, pp. 31--43, 1975.

\bibitem{KokenTuncel}
E.~{Köken} and E.~{Tuncel}, ``On minimum energy for robust {Gaussian} joint
  source-channel coding with a distortion-noise profile,'' in \emph{Proceedings
  of the IEEE International Symposium on Information Theory (ISIT)}, Aachen,
  Germany, 2017, pp. 1668--1672.

\bibitem{baniasadi2020minimum}
M.~Baniasadi and E.~Tuncel, ``Minimum energy analysis for robust {Gaussian}
  joint source--channel coding with a square-law profile,'' in
  \emph{Proceedings of the IEEE International Symposium on Information Theory
  and Its Applications (ISITA)}, 2020, pp. 51--55.

\bibitem{EnergyLimitedJSCC:Universal:Lev_Khina:Full}
\BIBentryALTinterwordspacing
O.~Lev and A.~Khina, ``Universal joint source--channel coding under an input
  energy constraint,'' Tech. Rep., Jan. 2021. [Online]. Available:
  \url{https://www.eng.tau.ac.il/~anatolyk/papers/journal/energy_limited_universal.pdf}
\BIBentrySTDinterwordspacing

\bibitem{Reznic}
Z.~Reznic, M.~Feder, and R.~Zamir, ``Distortion bounds for broadcasting with
  bandwidth expansion,'' \emph{IEEE Transactions on Information Theory},
  vol.~52, no.~8, pp. 3778--3788, Aug.~2006.

\bibitem{EnergyLimitedJSCC:Lev_Khina:ITW2021}
O.~Lev and A.~Khina, ``Energy-limited joint source--channel coding via analog
  pulse position modulation,'' in \emph{Proceedings of the IEEE Information
  Theory Workshop (ITW)}, Oct. 2021, accepted.

\bibitem{Q-function:approx:com1979}
P.~Borjesson and C.-E. Sundberg, ``Simple approximations of the error function
  {$Q(x)$} for communications applications,'' \emph{IEEE Transactions on
  Communications}, vol.~27, no.~3, pp. 639--643, Mar. 1979.

\bibitem{shepp1966radon}
L.~A. Shepp, ``Radon--{Nikodym} derivatives of {Gaussian} measures,'' \emph{The
  Annals of Mathematical Statistics}, pp. 321--354, 1966.

\end{thebibliography}
